\documentclass[twocolumn,amsmath,amssymb]{revtex4}

\def\gsim{ \lower .75ex \hbox{$\sim$} \llap{\raise .27ex \hbox{$>$}} }

\def\lsim{ \lower .75ex \hbox{$\sim$} \llap{\raise .27ex \hbox{$<$}} }

\def\bR{{{\mathbb R}}}
\def\ie{{\it i.e.}}
\def\IZ{\relax\ifmmode\mathchoice
{\hbox{\cmss Z\kern-.4em Z}}{\hbox{\cmss Z\kern-.4em Z}}
{\lower.9pt\hbox{\cmsss Z\kern-.4em Z}} {\lower1.2pt\hbox{\cmsss
Z\kern-.4em Z}}\else{\cmss Z\kern-.4em Z}\fi}
\def\IR{\relax{\rm I\kern-.18em R}}

\def\one{{\hbox{ 1\kern-.8mm l}}}

\newlength{\bredde}
\def\slash#1{\settowidth{\bredde}{$#1$}\ifmmode\,\raisebox{.15ex}{/}
\hspace*{-\bredde} #1\else$\,\raisebox{.15ex}{/}\hspace*{-\bredde}
#1$\fi}

\newsavebox{\zzzbar}
\sbox{\zzzbar}
  {\setlength{\unitlength}{0.9em}
  \begin{picture}(0.6,0.7)
  \thinlines
  \put(0,0){\line(1,0){0.6}}
  \put(0,0.75){\line(1,0){0.575}}
  \multiput(0,0)(0.0125,0.025){30}{\rule{0.3pt}{0.3pt}}
  \multiput(0.2,0)(0.0125,0.025){30}{\rule{0.3pt}{0.3pt}}
  \put(0,0.75){\line(0,-1){0.15}}
  \put(0.015,0.75){\line(0,-1){0.1}}
  \put(0.03,0.75){\line(0,-1){0.075}}
  \put(0.045,0.75){\line(0,-1){0.05}}
  \put(0.05,0.75){\line(0,-1){0.025}}
  \put(0.6,0){\line(0,1){0.15}}
  \put(0.585,0){\line(0,1){0.1}}
  \put(0.57,0){\line(0,1){0.075}}
  \put(0.555,0){\line(0,1){0.05}}
  \put(0.55,0){\line(0,1){0.025}}
  \end{picture}}

\def\Im{{\rm Im ~}}

\newcommand{\ena}{\end{eqnarray}}
\newcommand{\beqa}{\begin{eqnarray}}
\newcommand{\eeqa}{\end{eqnarray}}
\newcommand{\bea}{\begin{eqnarray}}
\newcommand{\eea}{\end{eqnarray}}

\newcommand{\be}{\begin{equation}}
\newcommand{\ee}{\end{equation}}

\newcommand{\Tr}{{\rm Tr}}
\usepackage{graphicx}

\def\im{{{\rm i}}}
\def\nn{\nonumber}
\def\ben{\begin{equation}}
\def\een{\end{equation}}

\def\bea{\begin{eqnarray}}
\def\eea{\end{eqnarray}}
\def\be{\begin{equation}}
\def\ee{\end{equation}}
\def\beq{\begin{eqnarray}}
\def\eeq{\end{eqnarray}}

\def\({\left (}
\def\){\right )}
\def\[{\left [}
\def\[{\right ]}

\def\ba{\begin{eqnarray}}
\def\ea{\end{eqnarray}}

\input amssym.def
\input amssym.tex

\usepackage{dcolumn}

\begin{document}

\title{Naturalness of CP Violation in the Standard Model}

\author{Gary W. Gibbons,$^1$ Steffen Gielen,$^1$ C. N. Pope$^{1,2}$ and Neil Turok$^{1,3}$}

\affiliation{
\vskip .5mm
{\it $^1$DAMTP, Centre for Mathematical Sciences, Cambridge University, Wilberforce Road, Cambridge, CB3 0WA, UK }
\vskip .5mm
{\it $^2$George P. \& Cynthia W. Mitchell Institute for Fundamental Physics and Astronomy,
\\ Texas A\&M University, College Station, TX 77843-4242, USA}
\vskip .5mm
{\it$^3$Perimeter Institute for Theoretical Physics, 31 Caroline St. N., Waterloo, Ontario, Canada N2L 2Y5}
}

\begin{abstract} 
We construct a natural measure on the space of CKM matrices in the standard model, assuming the fermion mass matrices are randomly selected from a distribution which incorporates the observed quark mass hierarchy. This measure allows us to assess the likelihood of Jarlskog's CP violation parameter $J$ taking its observed value $J \approx 3 \times 10^{-5}$. We find that the observed value, while well below the mathematically allowed maximum, is in fact typical once the observed quark masses are assumed. 
\end{abstract}


\maketitle


\section{Introduction}

Kobayashi and Maskawa's beautiful explanation for why CP violation is naturally expected in the standard model of particle physics, with three or more families \cite{km}, has been recently rewarded with a Nobel prize. Still, the {\it magnitude} of the observed CP violation remains unexplained, and this is a major task awaiting a fundamental theory. Even before we have such a theory, it is interesting to ask whether the observed CP violation is ``finely tuned," and therefore poses a major new puzzle, or whether it is in some sense ``typical," of what a fundamental framework might be expected to predict. 

According to Kobayashi and Maskawa, the observed CP violation originates in the Yukawa couplings between right- and left-handed quarks. These couplings are not entirely physical because they can be altered by a global redefinition of the basis of fermion fields. To resolve this ambiguity, Jarlskog has constructed a basis-independent measure of CP violation, called $J$. In this Letter, we construct a natural measure on the space of mass matrices, assuming the observed quark mass hierarchy but predicting the CKM mixing angles and phase. From this we infer the typically expected value for $J$. We find that the observed value of $J$, while far smaller than the maximal mathematically allowed value, is indeed very likely. Hence we conclude that, at least with the measure we define, there is no ``fine tuning" problem in this aspect of the standard model. 

We begin from the low energy effective Lagrangian describing standard model quarks. In the gauge basis, this contains mass terms 
\be
{\cal L}_{\rm mass} = - \overline{Q}_L \, m \,{Q}_R -\overline{Q'}_L \, m'  \,{Q'}_R +{\rm h.c.},
\label{eq1}
\ee
where $Q=(t,c,u)$, $Q'=(b,s,d)$ and $m$ and $m'$ are arbitrary $3\times 3$ complex matrices, with dimensions of mass. Following \cite{jarlskog1}, we define the dimensionless mass matrices $M=m/\Lambda$ and $M'=m'/\Lambda'$, where $\Lambda$ and $\Lambda'$ are mass scales which may be chosen for convenience. Via a redefinition of $Q_R$ and $Q'_R$, $M$ and $M'$ may be rendered Hermitian, with non-negative eigenvalues. They can be represented as follows
\be 
M=U^\dagger D U, \qquad M'= U'^\dagger D'U',
\label{eq2}
\ee 
where $D$ and $D'$ are real and diagonal, $D\equiv{\rm diag}(M_t,M_c,M_u)$ and $D'\equiv{\rm diag}(M_b,M_s,M_d)$. The final step is to redefine $U Q_L =Q_L^{\rm mass}$ and similarly for $Q_R, Q'_L,Q_R'$, removing $U$ and $U'$ from the mass terms. They now appear only in the charged current interaction terms in the Lagrangian, $J_{cc}^\mu = \overline{Q}_L^{\rm mass} \gamma^\mu V {Q'}_L^{\rm mass} +{\rm h.c.}$, with $V= U U'^\dagger$ being the Cabibbo-Kobayashi-Maskawa (CKM) mixing matrix.

The normalized mass eigenstates are only defined up to a phase. Jarlskog  defined a measure of CP violation which is invariant under the choice of these phases~\cite{jarlskog1,jarlskog2},
\ben
J= -\im \det[M,M^\prime \bigr]/(2 TB)\, ,
\een
where
\bea
& T=(M_t-M_u)(M_t-M_c)(M_c-M_u)\,,\nn
\\ & B=(M_b-M_d)(M _b-M_s)(M_s-M _d)\,.
\eea
In terms of the CKM matrix $V$,
\ben
J= \Im \bigl (V_{11} \,V_{22}\, V_{12}^* \, V_{21}^* \bigr )\,,
\een
and $|J|$ has an elegant geometrical interpretation as twice the area of the so-called unitarity triangle.

Mathematically, (\ref{eq2}) represents the space of mass matrices $M$ (with three distinct eigenvalues) as $(U(1)^2\times \frak{S}_3)\backslash SU(3)\times \bR^{3}$, for generic $D$, because left-multiplication of $U$ by a diagonal phase matrix $P \in SU(3)$, {\it i.e.}, an element of $U(1)^2$, does not alter $M$. To avoid overcounting, we must in addition factor out by elements of $SU(3)$which permute the masses, as we shall detail below ($\frak{S}_3$ is the symmetric group on 3 elements).  Because only the product $U U'^\dagger$ appears in physical quantities, the two $SU(3)$'s are redundant and we may identify them with a diagonal $SU(3)$ subgroup. Thus the final physical space of angles and phases encoded in $V$ may be thought of as the {\sl double coset} $(U(1)^2\times \frak{S}_3)\backslash SU(3)/(U(1)^2\times \frak{S}_3)$. Our goal is to construct a natural measure on this space which incorporates the observed values of the quark masses.

The standard measure on the space of Hermitian matrices, $D M$, will be detailed shortly. Since the space of eigenvalues is not compact, we must also include a weight function which makes the integral over the eigenvalues converge. We choose the weight function to distinguish the three up-type quarks and give them appropriate expected masses, and similarly for the three down-type quarks.
The simplest assumption is that the measure factorizes,
\be 
DM f\left(\Tr(M^2 A)\right) D M' f\left(\Tr((M')^2 A')\right),
\label{eq3}
\ee 
where $A$ and $A'$ are Hermitian and positive definite, and we shall further assume $[A,A']=0$.

Since physical quantities are invariant under $U\rightarrow U W$, $U'\rightarrow U' W$, without loss of generality we can choose $A$ and $A'$ to be diagonal. For simplicity and ease of technical calculations, we shall choose the function $f$ in (\ref{eq3}) to be a decaying exponential so $M$ and $M'$ are governed by Gaussian distributions.  Our proposal is to fit the diagonal matrices $A$ and $A'$ to the observed quark masses, and then employ the resulting distribution given by (\ref{eq3}) to infer a probability measure on $J$. 

The introduction of the matrices $A$ and $A'$ means that the invariance of the measure $DM\,DM'$ under separate conjugation of $M$ and $M'$ by arbitrary elements of $U(3)$, \ie, under the action of $U(3)\times U(3)$, is broken down to the action of the diagonal subgroup $U(1)^2 \times U(1)^2$ which commutes with $A$ and $A'$. We find that this symmetry breaking is necessary to obtain a distribution that reproduces different expectation values for squared quark masses.

Full details of the calculations presented in this Letter, as well as a more general discussion of possible measures on the space of CKM matrices, may be found in the separate publication \cite{longpaper}.

\section{Calculating the Measure}

The natural line element on the space of Hermitian matrices is 
\bea
ds^2 & = &\Tr(dM\, dM)
\\& = &\Tr\left(dD\, dD\right)+2\Tr\left(\left(dU\,U^{\dagger}\,D\right)^2-\left(dU\,U^{\dagger}\right)^2 D^2 \right)\nn
\eea
which is invariant under conjugation under $U(3)$. Our parametrization for the matrix $U$, an element of the space $U(1)^2\backslash SU(3)$, is
\ben
U=W\, T_R\,,
\een
where
\ben
W=\begin{pmatrix} c_y c_z & c_y s_z & e^{-\im w} s_y\cr 
    -c_x s_z - e^{\im w} s_x s_y c_z & c_x c_z - e^{\im w} s_x s_y s_z&
    s_x c_y\cr 
   s_x s_z - e^{\im w} c_x s_y c_z & - s_x c_z - e^{\im w} c_x s_y s_z &
    c_x c_y \end{pmatrix} \,,
\een
and
\ben
T_R=\hbox{diag}\Big(e^{\im r + \im t}, e^{\im r - \im t},
    e^{-2\im r}\Big)\,,
\een
where $s_x=\sin x$, $c_x=\cos x$, etc., and the ranges of the coordinates on $U(1)^2\backslash SU(3)$ are
\ben
x,y,z\in \left[0,\frac{\pi}{2}\right],\quad w,r,t\in \left[ 0, 2\pi \right]\,.
\een
In this parametrization, the Jarlskog invariant $J$ is given by
\ben
J = \frac{1}{4} \sin2x\,  \sin2z\, \sin y\, \cos^2 y\, \sin w\,,
\een
which has a maximal value of $\frac{1}{6\sqrt{3}}\approx 0.096$. 

We obtain a Riemannian measure
\ben
DM:=\prod_{i < j}(D_i-D_j)^2 \,dD_1\,dD_2\,dD_3\, DU\,,
\label{hermit}
\een
with
\ben
DU:=\sin 2x\,\cos^3 y\,\sin y\,\sin 2z\,dx\,dy\,dz\,dw\,dr\,dt\,,
\een
on the space of Hermitian $3\times 3$ matrices. We allow these to have arbitrary eigenvalues. (In a fermionic mass term, the sign of the mass has no physical significance, and only $m^2$ enters in physical quantities.) Note that the subspace of matrices with coinciding eigenvalues has measure zero and can be ignored in the following discussion.

Each Hermitian matrix with three distinct eigenvalues is associated with six different elements of $\bR^3 \times U(1)^2\backslash SU(3)$, related by the action of the discrete group $\frak{S}_3$:
\ben
U^{\dagger}DU=(U^{\dagger}P^{-1})PDP^{-1}(PU)=:\tilde{U}^{\dagger}\tilde{D}\tilde{U}\,, \; P\in \frak{S}_3\,,
\een
where $\frak{S}_3$ is the symmetric group of degree 3 which permutes the canonical basis vectors of $\bR^3$. We need to consider the space $ (U(1)^2 \times \frak{S}_3)\backslash SU(3)\times \bR^3$ instead, restricting the coordinates on $U(1)^2\backslash SU(3)$ in order to pick one of the six matrices related by the $\frak{S}_3$ action. This can be achieved by demanding that the elements of the third column have ascending absolute value,
\ben
0\le y \le \arctan (\sin x)\,,\quad 0\le x\le \frac{\pi}{4}\,.
\een

Since there are two mass matrices and we need to average over two copies of the space of $3\times 3$ Hermitian matrices, an integral of a quantity such as $J^2$ becomes
\ben
\langle J^2 \rangle = \frac{\int DM \, DM'\, e^{-\Tr(M^2 A)-\Tr((M')^2 A')} \,J^2(M,M')}{\int DM \, DM'\, e^{-\Tr(M^2 A)-\Tr((M')^2 A')}}\,,
\label{jintegral}
\een
and $J$ really depends only on $V=UU'^{\dagger}$. We choose $J^2$ since all odd powers of $J$ average to zero. We use the same parametrization for $M'$ as for $M$, parametrizing $U'$ by $x',y',z',$ etc.

It should be clear from (\ref{jintegral}) that multiplying $A$ (or $A'$) by a constant is the same as rescaling the eigenvalues $D_i$ (or $D_i'$), and so amounts to a rescaling of the mass scales $\Lambda$ and $\Lambda'$. We can therefore, without any loss of generality, choose
\ben
A=\left(\begin{matrix}
1 & 0 & 0 \\ 0 & 1/\mu_c^2 & 0 \\ 0 & 0 & 1/\mu_u^2
\end{matrix}\right),\quad
A'=\left(\begin{matrix}
1 & 0 & 0 \\ 0 & 1/\mu_s^2 & 0 \\ 0 & 0 & 1/\mu_d^2
\end{matrix}\right)\, ,
\label{amatrices}
\een
where $\mu_c, \mu_u, \mu_s$ and $\mu_d$ are dimensionless parameters that we are free to choose so as to reproduce the observed squared quark masses as expectation values $\langle D_1^2 \rangle$ etc. We expect these to be of the same order of magnitude as the corresponding quark masses, expressed in units where $\Lambda=m_t$ and $\Lambda'=m_b$. 

In the case where $A$ and $A'$ are proportional to the identity, the exponential factor becomes $\exp(-\alpha\Tr(D^2)-\beta\Tr((D')^2))$, and the integral (\ref{jintegral}) splits into a product of an integral over the eigenvalues, and an integral of $J^2$ over $((U(1)^2\times\frak{S}_3)\backslash SU(3))^2$. The distribution of $J^2$ is then independent of the quark masses, and one would expect $J^2$ to be much larger than observed. A generic $A$ couples the diagonal $D$ to $U$ which is an element of $((U(1)^2\times\frak{S}_3)\backslash SU(3))^2$, and allows the quark masses to have an effect on the distribution of $J$.

\section{Results}

We need to use approximations to evaluate the integral (\ref{jintegral}), as the expression for $J$ in terms of coordinates on $((U(1)^2\times \frak{S}_3)\backslash SU(3))^2$ is too complicated to be given explicitly. However, since
\ben
\Tr(M^2 A)=\Tr(D^2 U A U^{\dagger})=\frac{D_1^2}{\mu_u^2} \sin^2 y+\ldots
\een
and
\ben
\Tr((M')^2 A')=\Tr((D')^2 U' A' (U')^{\dagger})=\frac{(D'_1)^2}{\mu_d^2} \sin^2 y'+\ldots
\een
with all remaining terms being non-negative, and $\mu_u\ll 1$ and $\mu_d\ll 1$, the integrand is negligibly small unless $y\approx 0$ and $y'\approx 0$. We can therefore do the integrations over $y$ and $y'$, using $\sin y\approx y$ with the remaining part of the integrand taken at $y=y'=0$. The result is now independent of $w$ and $w'$, and we can integrate the expression for $J^2$ over $r,r',t,$ and $t'$. Then the integrals over both copies of $\bR^3$ can be done analytically, but one is left with a four-dimensional integral over $x,z,x'$ and $z'$ which, in general, can only be evaluated numerically.

Expectation values for squared masses, however, take the relatively simple form
\ben
\langle D_1^2 \rangle \approx \frac{\int d^3D \, d^2 x\, D_1^2  \left(e^{-\Tr(D^2 UAU^{\dagger})}\right)\big|_{y=0}}{\int d^3D\,  d^2x \,  \left(e^{-\Tr(D^2 UAU^{\dagger})}\right)\big|_{y=0}}\,,
\een
where
\ben
\int d^3 D \,d^2 x:=\int_{\bR^3} \prod_k dD_k \prod_{i < j}(D_i-D_j)^2\int\limits_0^{\pi/4} dx \int\limits_0^{\pi/2} dz\,s_{2x} s_{2z}\,.
\een
Using the fact that these integrals are dominated by small $x$ and $z$, we find \cite{longpaper} the analytical approximations
\ben
\langle D_1^2 \rangle \approx \frac{3}{2}\,,\quad \langle D_2^2 \rangle \approx \frac{\mu_c^2}{2}\,,\quad \langle D_3^2 \rangle \approx \frac{\mu_u^2}{2}\,,
\label{ean}
\een
and similarly for $(D'_i)^2$.

We use these approximate analytical results as a consistency check for more accurate numerical calculations of expectation values of squared quark masses. The mass scales $\Lambda$ and $\Lambda'$ are fixed by setting $\langle D_1^2 \rangle=(m_t/\Lambda)^2$ and $\langle (D'_1)^2 \rangle=(m_b/\Lambda')^2$. By comparing the results obtained by numerical integration with the values we want to reproduce, we can then fix the parameters $\mu_c,\mu_u,\mu_s$ and $\mu_d$.

Due to the dependence of quark masses on the energy scale, described by the renormalization group, there is some ambiguity in what is meant by the ``quark masses" we want to reproduce. Following \cite{rosner}, for example, we take all the quark masses evolved to the scale of the $Z$ boson mass.
We use the central values \cite{massref}
\bea
(m_u,m_c,m_t) & = & (1.27\;{\rm MeV},\; 619\;{\rm MeV},\;171.7\;{\rm GeV})\,,\nn
\\ (m_d,m_s,m_b) & = &(2.9\;{\rm MeV},\; 55\;{\rm MeV},\;2.89\;{\rm GeV})\,.
\label{cvalues}
\eea

Comparing our numerical results with (\ref{cvalues}), we fix the parameters to
\bea
\mu_c^2 = 3\left(\frac{m_c}{m_t}\right)^2\,,\quad \mu_u^2 = 3\left(\frac{m_u}{m_t}\right)^2\,,\nn
\\ \mu_s^2 = \frac{3}{2}\left(\frac{m_s}{m_b}\right)^2\,,\quad \mu_d^2 = \frac{12}{5}\left(\frac{m_d}{m_b}\right)^2\,.
\label{muvalues}
\eea
to reproduce the corresponding expectation values for squared eigenvalues of the mass matrices.
(The analytical approximation (\ref{ean}) is less accurate for the $d,s,b$ quarks since the mass hierarchy is milder, so the numerical factors given here are numerical fits.) 

We can obtain an analytical expression for the expectation value of $J^2$ because the integral (\ref{jintegral}) is dominated by the region of small $y, y',$ and $z$, and therefore to a first approximation we only need $J^2$ at $y=y'=z=0$. Averaging over $r,t,r'$ and $t'$ gives a factor of 1/2, and therefore we can use
\ben
J^2_{{\rm small}\;y,y',z}:=\frac{1}{2}\sin^2 x\cos^2 x\sin^2 x'\cos^2 x'\cos^2 z'\sin^4 z'
\een
for our calculations. We obtain 
\ben
\langle J^2_{{\rm small}\;y,y',z}\rangle \approx \frac{\frac{4}{\sqrt{15}}m_u\,m_d\,m_b}{m_c \,m_s^2\left(1+\sqrt{\frac{2}{3}}\frac{m_b}{m_s}\right)^4}\, t_1 \, t_2\,,
\label{japprox}
\een
with
\bea
& t_1=\arctan\left(\frac{m_c}{2m_u}\right)-\frac{2m_u}{m_c}\,,\nn
\\& t_2=\arctan\left(\sqrt{\frac{5}{32}}\frac{m_s}{m_d}\right)-\sqrt{\frac{32}{5}}\frac{m_d}{m_s}\,,
\eea
where the different prefactors arise from the numerical factors given in (\ref{muvalues}). Note that the top quark mass does not appear in this leading approximation.

Numerical evaluation of $\langle J^2 \rangle$ (taken at $y=y'=0$, but integrated over all $z$) gives
\ben 
\Delta J = \sqrt{\langle J^2 \rangle} \approx 7.27 \times 10^{-5}. 
\een
Therefore the observed value $J\approx 3.08\times 10^{-5}$ \cite{particledb} is about 0.42 standard deviations away from zero in our distribution, and cannot be viewed as being finely tuned.

To test the sensitivity of our results to experimental uncertainties in the quark masses, we take their values at the upper or lower limits given in \cite{massref} and obtain the corresponding highest and lowest values for $\langle J^2\rangle$. We find the bounds
\ben
 4.31 \times 10^{-5} \le \Delta J \le 1.23 \times 10^{-4}\,.
\een

By identifying the dominant regions of the integrals appearing in (\ref{jintegral}), we are also able to give estimates \cite{longpaper} of the expected magnitudes of the components $V_{td},\;V_{ts},\;V_{cd}$ of the CKM matrix \cite{particledb}, finding
\ben
|V_{td}| \approx \mu_d \approx 0.002\,,\; |V_{ts}| \approx \mu_s \approx 0.02\,,\;|V_{cd}| \approx \frac{\mu_d}{\mu_s} \approx 0.07\,.
\een
These very rough estimates are within factors of a few of the observed values, and reproduce the correct ordering, lending further support to the approach taken here.

\section{Conclusions and Outlook}

In this Letter, we have shown that by assuming the observed hierarchy in quark masses in a Gaussian distribution over the space of mass matrices one obtains expectation values for $J^2$ close to the observed value. This statistical observation indicates that the mechanism responsible the quark mass hierarchy might also explain why the observed value for $|J|$ is so small, with no additional fine tuning. In strong contrast, a geometrical measure on the space of CKM matrices alone, which does not take quark masses into account, makes the observed value of $J$ appear highly unlikely~\cite{longpaper}. 

These results suggest that the choice of CKM angles is closely linked to the observed mass hierarchy. Broadly similar conclusions have been reached before~\cite{ddr}, but we believe our work represents a significant improvement. First, the measure we use is invariant under unitary rotations of the quark fields $Q_R$ and $Q_R'$ whereas the measure in \cite{ddr} is not and hence depends on arbitrary conventions. Second, we have input the observed quark masses and obtained fairly tight predictions for $\langle J^2 \rangle$, still consistent with observation. In contrast, \cite{ddr} assumed power law distributions for the magnitudes of the elements of $M$ and $M'$ and obtained the much looser prediction $10^{-8} < |J| < 10^{-4}$ at one sigma. Finally, we have obtained analytical approximations whereas their work was entirely numerical.  

Our analysis also applies to the case of massive neutrinos, where the predictions will conceivably be tested by future experiments. In the standard theory, the Maki-Nakagawa-Sakata matrix \cite{mns} which appears is naturally an element of $U(1)^2\backslash SU(3)$, since only phasing of the lepton charge eigenstates, but not the neutrino mass eigenstates (which are assumed Majorana) is possible. Since the right phases do not play any role in neutrino oscillations, and the relevant $J$ is independent of these phases, the calculations are identical to the ones presented here, although with the appropriate values of the $\mu$ parameters appearing in $A$ and $A'$.
In the see-saw mechanism one adds very heavy right-handed neutrinos, and the most general mixing matrix would be an element of $U(1)^5 \backslash SU(6)$. This is naturally a K\"ahler manifold, and the measure induced by the K\"ahler metric can be obtained from the analysis in \cite{picken}. We leave a detailed treatment of this case to future work.

Finally, one could analyze the effects of a fourth generation of quarks on CP violation, by repeating the calculations for $4\times 4$ Hermitian matrices. If this generalization spoils the agreement with the observed $J$, one might obtain interesting lower bounds on the masses of a hypothetical extra generation of quarks.

SG acknowledges funding from EPSRC and Trinity College, Cambridge. We thank Ben Allanach for helpful conversations and Malcolm Perry for suggesting the effect of a fourth generation. This research was also supported by Perimeter Institute for Theoretical Physics.

\nopagebreak

\end{document}